\documentclass[aps,prl,amsmath,amssymb]{revtex4}
\usepackage{graphicx}

\begin{document}

\title{Exciton Ionization, Franz-Keldysh and Stark Effects in Carbon Nanotubes}

\author{Vasili Perebeinos$*$ and Phaedon Avouris}
\affiliation{IBM Research Division, T. J. Watson Research Center,
Yorktown Heights, New York 10598 \\
{\it\bf\Large Nano Letters 2007}}

\date{\today}

\begin{abstract}
We calculate the optical properties of carbon nanotubes in an
external static electric field directed along the tube axis. We
predict strong Franz-Keldysh oscillations in the first and second
band-to-band absorption peaks, quadratic Stark effect of the first
two excitons, and the field dependence of the bound exciton
ionization rate for a wide range of tube chiralities. We find that
the phonon assisted mechanism dominates the dissociation rate in
electro-optical devices due to the hot optical phonons. We predict a
quadratic dependence of the Sommerfeld factor on the electric field
and its increase up to 2000\% at the critical field of the full
exciton dissociation.
\end{abstract}

\maketitle

Semiconducting carbon nanotubes are direct bandgap materials which
have attracted much attention recently for nanophotonic applications
\cite{Avouris}. Almost fifty years ago Franz and Keldysh predicted
that a static electric field would modify the linear optical
properties of the  3D semiconductors near their absorption edge
\cite{Franz,Keldysh}. They showed that the absorption coefficient
decays exponentially for photons below the bandgap and shows
oscillations for energies above the bandgap.  The interest in
electroabsorption was revived about three decades latter after
discovery of the Quantum-Confined Stark Effect in 2D quantum well
structures \cite{Miller1, Miller2}. Large Stark shifts were observed
in fields directed perpendicular to the 2D planes. In  1D  carbon
nanotubes, excitons were predicted to have large binding energies
\cite{Ando} and to dominate the absorption spectra
\cite{Perebeinos1,Spataru}, a fact which was verified experimentally
by two-photon spectroscopy \cite{Wang,Maultzsch} and from the
observation of the phonon sidebands in photoconductivity spectra
\cite{Freitag}.  The exciton binding energies in carbon nanotubes
have interesting scaling properties \cite{Perebeinos1} and they can
be as small as those in 2D structures and as large as 30\% of their
bandgap depending on both the nanotube structure and the
environment.

It has been long recognized that excitonic effects enhance
significantly the electroabsorption signal \cite{Dow}. An electric
field leads to several modifications of the absorption spectrum: 1)
modulation of the absorption coefficient; 2) growth of the
band-to-band absorption spectral weight; 3) shift of the absorption
peak, known as the Stark effect; and 4) dissociation of the bound
exciton. In bulk  3D semiconductors the binding energy is small and
most of the theoretical and experimental focus has been on the field
induced absorption in the region below the bandgap and on the
quantum confined Stark effect in 2D structures \cite{Haug}. In
carbon nanotubes, the binding energy is large and the oscillator
strength of the higher lying Rydberg states is infinitesimal, so
that relatively large changes in the absorption at the first
excitonic peak and the first band-to-band absorption are expected.

Here we explore the field induced changes in the absorption spectra
of nanotubes for a large range of photon energies as a function of
tube chirality and dielectric environment. In opto-electronic carbon
nanotube devices, excitons can be produced by electrom-hole
recombination or by impact excitation by hot carriers
\cite{Chen,Martel,Perebeinos5}. Therefore, we also address here the
problem of the bound exciton dynamics produced in the impact
excitation process in an electric field and in the presence of the
hot optical phonons. We use a Bethe-Salpeter equation solution for
an exciton in the static electric field directed along the tube
axis.  Similar to Ref.~\cite{Perebeinos1} we employ periodic
boundary conditions for the unit cell of length $L$ (typically
$L\approx450$ nm) where the electron-hole interaction is computed by
including both the direct and exchange Coulomb terms
\cite{Rohlfing}. The nanotube diameter dependence of the exciton
binding energies obtained by two-photon fluorescence excitation
spectroscopy \cite{Dukovic} agrees very well with our model
calculations \cite{Perebeinos1} if we choose $\varepsilon=3.3$, a
value that we use in the rest of the paper, unless stated otherwise.
The external electric field potential is added to the total
Hamiltonian in the Bethe-Salpeter equation. The potential  for the
relative motion of the exciton has been smoothed out to avoid
numerical instabilities:
\begin{eqnarray}
U_F(\vec{r}_e-\vec{r}_h)&=&eF\left(z_e-z_h\right) \tanh\left(k_0\left(\frac{1}{4}-\frac{\left(z_e-z_h\right)^2}{L^2}\right)\right)
\label{eq1}
\end{eqnarray}
 such that the potential is zero at the unit-cell boundaries at
$\vert z_e-z_h\vert =\pm L/2$, while at small $\vert z_e-z_h\vert
\ll L/2$ the hyperbolic tangent equals to unity if we choose the
smoothing parameter $k_0$ to be $\ge30$. The absorption spectra are
calculated as in Ref.\cite{Perebeinos1} from the $q=0$ excitonic
wavefunctions.

\begin{figure}[h!]
\includegraphics[height=2.4in]{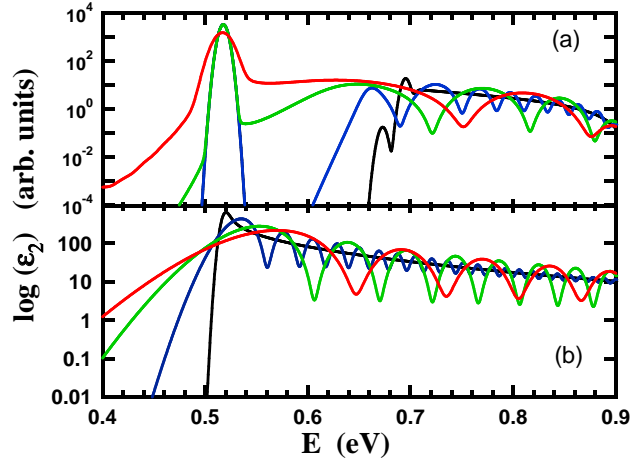}
\caption{\label{FigSpect} Absorption spectra in (16, 8) tube $d=1.7$
nm in applied electric fields $F$ in V/$\mu$m, $F=$ 0.0 (black), 2.0
(blue), 6.0 (green), 10.0 (red) with excitons (a) and for the
band-to-band absorption (b) with no excitonic effects.  The GW
corrections are modeled by the scissor approximation with a scissor
value equal to the first exciton binding energy  of 194 meV, such
that the onsets of zero field spectra in (a) and (b) coincide.}
\end{figure}

The absorption spectra at different fields are shown in
Fig.~\ref{FigSpect}a. At zero field, there is no absorption in the
energy range between the first $E_{11}$ exciton  at 0.52 eV and the
onset of the first band-to-band (free electron-hole continuum)
absorption  $\Delta_{11}$ at 0.71 eV. The higher energy Rydberg
states have an infinitesimal spectral weight in 1D, except for the
highest energy Rydberg state at 0.70 eV, whose energy is in
resonance with the onset of the first band-to-band absorption. The
interaction between this excitonic state and the band-to-band
continuum leads to a spectral weight transfer to the Rydberg
exciton, which appears as a peak in the absorption spectra just
below the onset of the first band-to-band absorption in
Fig.~\ref{FigSpect}a.

 At a finite electric field, the band-to-band absorption is
modulated and grows in intensity, as shown in Fig.~\ref{FigSpect}a.
Similar to the free electron-hole absorption, in the absence of
excitonic effects, shown in Fig.~\ref{FigSpect}b, the band-to-band
absorption is modulated  with a period proportional to the field
strength. However, the modifications of the band-to-band absorption
in the forbidden region (i. e. below $\Delta_{11}$) in the absence
of excitonic effects, and correspondingly the absorption below
$\Delta_{11}$ and above the excitonic $E_{11}$ peak in the presence
of excitons, are very different. In the former case, the free
particle absorption decays exponentially in the bandgap region as
shown in Fig.~\ref{FigSpect}b, while in the excitonic picture an
additional peak develops in the forbidden region, as shown in
Fig.~\ref{FigSpect}a. As the field is increased, this peak grows in
intensity, moves deeper in the forbidden region, and eventually
merges with the $E_{11}$ exciton absorption peak. We define as the
exciton dissociation field $F_c$ the field at which the first
Franz-Keldysh peak coincides with the $E_{11}$ exciton energy. At
the same time, the ``transparent'' region between the first two
oscillatory peaks broadens and the first absorption minimum moves
towards higher energies above the zero-field bandgap edge
$\Delta_{11}$. Above the second $E_{22}$ exciton (not shown in
Fig.~\ref{FigSpect}a), the second band-to-band absorption above the
second bandgap $\Delta_{22}$ also shows Franz-Keldysh modulations.
However, the absorption coefficient does not go to zero at energies
below $\Delta_{22}$, because the first band contributes to the total
absorption in that region.

\begin{figure}[h!]
\includegraphics[height=2.4in]{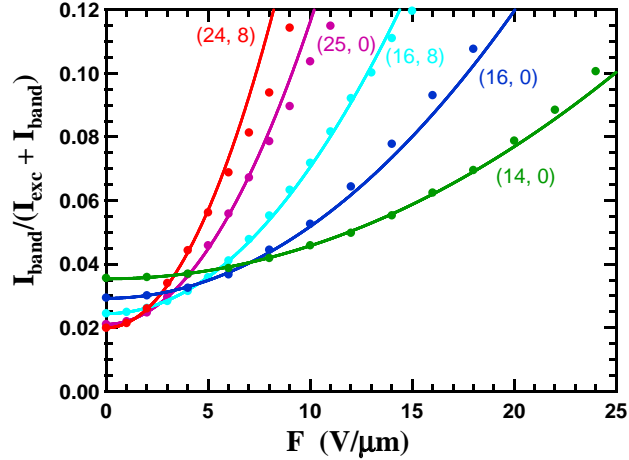}
\caption{\label{FigFree} The relative band-to-band absorption
spectral weight as a function of field in (14, 0) tube with $d=1.1$
nm (green circles), (16, 0) $d=1.3$ nm (blue circles), (16, 8)
$d=1.7$ nm (cyan circles), (25, 0) $d=2.0$ nm (magenta circles), and
(24, 8) $d=2.3$ nm (red circles). The solid curves are the best fits
to Eq.~(\ref{eq2}) with $A=0.009$ and  $(d_0, \kappa_s)$ from
table~\protect{\ref{tab1}}.}
\end{figure}

\begin{table}
\begin{ruledtabular}
\begin{tabular}{cccccccc}
\hline \hline
   (n, m)    & d  & $E_b$ &  $F_{0}$ & $d_0$ & $\kappa_s$ & $\kappa_b$ & $\alpha$ \\
\hline
    (13, 0)&  1.03    & 281 & 169  & 0.026  & 9.1  & 4.2 & 6.3 \\
    (14, 0)&  1.11    & 288 & 198   & 0.029  & 7.0  & 3.0 & {\it 9.9}\\
    (12, 4)&  1.15    & 277 & 179   & 0.029  & 7.2  & 3.2 & {\it 7.5}\\
    (16, 0)&  1.27    & 236 & 119  & 0.026  & 7.8  & 4.0 & 5.1 \\
    (17, 0)&  1.35    & 241 & 136  & 0.028  & 6.4  & 3.0 & 5.0 \\
    (19, 0)&  1.51    & 204 &  89  & 0.025  & 7.1  & 3.7 & 3.5 \\
    (20, 0)&  1.59    & 208 &  99  & 0.027  & 6.0  & 3.1 & 4.5 \\
    (16, 8)&  1.68    & 194 &  84  & 0.026  & 6.1  & 3.2 & 4.1 \\
    (22, 0)&  1.75    & 179 &  69  & 0.025  & 6.6  & {\it 3.6} & 3.3 \\
    (23, 0)&  1.83    & 182 &  75  & 0.026  & 5.8  & 3.1 & 3.9 \\
    (25, 0)&  1.99    & 160 &  54  & {\it 0.024}  & {\it 6.2}  & 3.6 & 3.8 \\
    (26, 0)&  2.06    & 162 &  59  & 0.025  & 5.5  & {\it 3.1} & 3.7 \\
    (24, 8)&  2.29    & 142 &  43  & {\it 0.025}  &{\it 5.7}  & 3.5 & 4.2 \\
\hline \hline
   \end{tabular}
\end{ruledtabular}
\caption{\label{tab1} Tube indices, diameters in nm, exciton binding
energy in meV, and $F_0$ fields in V/$\mu$m in semiconducting carbon
nanotubes studied here. The rest of the columns give the best fits
for $d_0$ in nm, $\kappa_s$, $\kappa_b$, and $\alpha$ in
Eq.~\protect{\ref{eq2}}, \protect{\ref{eq3}}, \protect{\ref{eq4}}.
In italic are shown fit parameters for an averaged fit error larger
than 8\%. Parameters $A=0.009$ and $\beta=1.74$ are kept fixed in
the fits.}
\end{table}

The spectral weight growth of the band-to-band absorption with the
electric field is due to the bound exciton wavefunction mixing with
the first band-to-band continuum, which leads to spectral weight
transfer from the excitonic peak to the latter. As a result, the
overall oscillator strength of the first band-to-band absorption
grows as much as 400\% before the excitonic peak starts to overlap
with the tail of the first Franz-Keldysh oscillatory peak. This
corresponds to the field strength of about 1/3 of the critical field
$F_c$ required for the full exciton dissociation. Upon exciton
dissociation the oscillator strength is being transferred to the
band-to-band absorption, which is increased from its zero-field
value by 10-20 times. The field dependence of the {\it Sommerfeld
factor}, defined here as the fraction of the band-to-band oscillator
strength, is shown in Fig.~\ref{FigFree} and it can be well fitted
by the following equation:
\begin{eqnarray}
\frac{I_{band}}{I_{band}+I_{exc}}=A+\frac{d_0}{d}+\kappa_s\frac{\left(edF\right)^2}{E_{b}^2}
\label{eq2}
\end{eqnarray}
where $A$, $d_0$, and $\kappa_s$ are fit parameters, $d$ is the tube
diameter, $F$ is the field strength, $e$ - electron charge, and
$E_b$ is the first exciton binding energy. The field dependence in
Eq.~(\ref{eq2}) is motivated by first order perturbation theory for
the bound exciton wavefunction mixed by the field with the first
band-to-band states separated in energy roughly by the exciton
binding energy. The matrix element for the coupling is proportional
to the field and the exciton size, which in turn is proportional to
the tube diameter \cite{Perebeinos1}. The best fit for all 13 tubes,
studied here, is achieved with $A=0.009$, $d_0=0.026$ nm, and
$\kappa_s=6.4$. The exact values used for each individual fit for
all 13 tubes are given in table \ref{tab1}. In agreement with our
predictions the exciton spectral weight reduction has been found to
scale quadratically with the applied electric field in recent
electroabsorption measurements\cite{Kennedy}.

\begin{figure}[h!]
\includegraphics[height=2.4in]{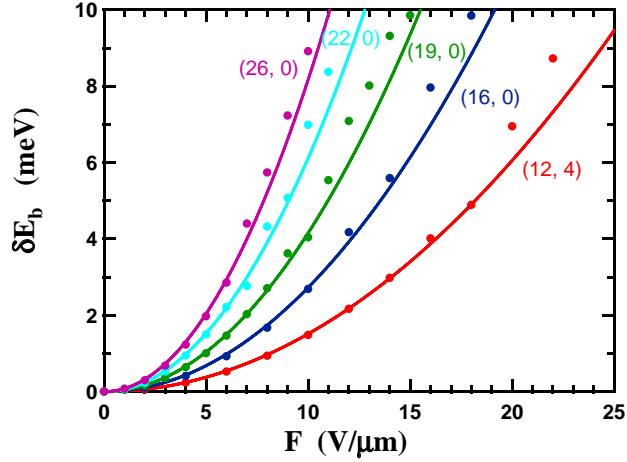}
\caption{\label{FigBind} Exciton binding energy dependence on
electric field in (12, 4) tube with $d=1.15$ nm (red circles), (16,
0) $d=1.3$ nm (blue circles), (19, 0) $d=1.5$ nm (green circles),
(22, 0) $d=1.7$ nm (cyan circles), and (26, 0) $d=2.1$ nm (magenta
circles). The solid curves are the best fits to Eq.~(\ref{eq2}) with
$\kappa_b$ from table~\protect{\ref{tab1}}.}
\end{figure}

The first exciton binding energy increases quadratically with the
field, leading to a red-shift of the $E_{11}$ absorption peak.
Despite the fact that the first exciton is nearly degenerate with
the lower energy dark state
\cite{Perebeinos1,Zhao,Perebeinos2,Spataru2}, we do not find enough
mixing between these two states to lead to an observable linear
Stark component. The linear Stark effect would be expected only if
the off-diagonal coupling strength between the dark and the bright
excitons is larger than its energy splitting. In zig-zag tubes the
mixing is identically zero and it is negligibly small in chiral
tubes. This is due to the different spatial symmetries of these two
states, which lead to a negligible dipole moment between them, as
was pointed out in Ref~\cite{Barros}.  The field dependence of the
binding energy is shown in Fig.~\ref{FigBind}. It is quadratic at
low fields, but it deviates from the parabola at higher fields. The
quadratic part of the Stark effect can be well described by the
following equation:
\begin{eqnarray}
\delta E_{b}=\kappa_b\frac{\left(edF\right)^2}{E_{b}} \label{eq3}
\end{eqnarray}
 whose form is similar to the second order perturbation theory for
the bound exciton energy and the exact treatment of the Stark effect
in hydrogen atom \cite{Alexander}. A quadratic shift was also
observed in the quantum confined Stark effect\cite{Miller1,Miller3}.
The best fit for all 13 tubes studied here is achieved with
$\kappa_b=3.4$. Similar arguments apply to the field dependence of
the lower energy dark state. As a result, we find the splitting
between the bright and the dark states to be nearly independent of
the field strength.

The second $E_{22}$ exciton binding energy has a similar quadratic
field dependence suggesting a red shift of the absorption peak, but
the magnitude of the shift is typically smaller by a factor of
three. The $E_{22}$ exciton can be coupled by the field to states
both higher and lower in energy, which would lead to opposite
contributions to the sign of the Stark shift.  In principle, the
sign of the net effect may become reversed, depending on the
strength of the couplings to the first and second band states. For
$\varepsilon=3.3$, we find $E_{22}$ to always red-shift (increase of
the binding energy) with field, but the net effect is reduced in
magnitude. However, when we reduce the dielectric constant to
$\varepsilon=2.0$, the second exciton binding energy increases, the
$E_{22}$ resonance comes close to the first band-to-band absorption
peak, and the coupling to the latter dominates leading to an overall
$E_{22}$ blue shift. At even higher fields of about $F\ge10$
V/$\mu$m and $\varepsilon=2.0$, we find that the sign of the second
exciton Stark shift suddenly reverses and we obtain a red-shift.
This nonperturbative result can be explained by the Franz-Keldysh
modulation of the band-to-band states.

\begin{figure}[h!]
\includegraphics[height=2.4in]{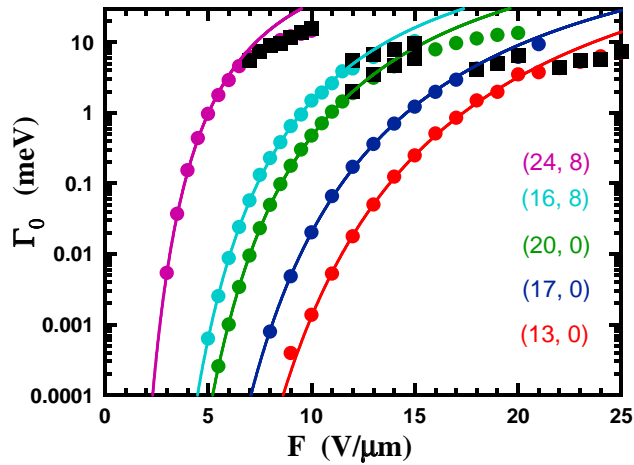}
\caption{\label{FigGamma} Exciton dissociation rate as a function of
field in (13, 0) tube with $d=1.0$ nm (red), (17, 0) $d=1.35$ nm
(blue), (20, 0) $d=1.6$ nm (green), (16, 8) $d=1.7$ nm (cyan), and
(24, 8) $d=2.3$ nm (magenta). The black squares are the linewidths
from the absorption fit to the Fano lineshape. The solid curves are
the best fit to Eq.~(\ref{eq1}) with $\beta=1.74$ and $\alpha$ from
table~\protect{\ref{tab1}}.}
\end{figure}

\begin{figure}[h!]
\includegraphics[height=2.4in]{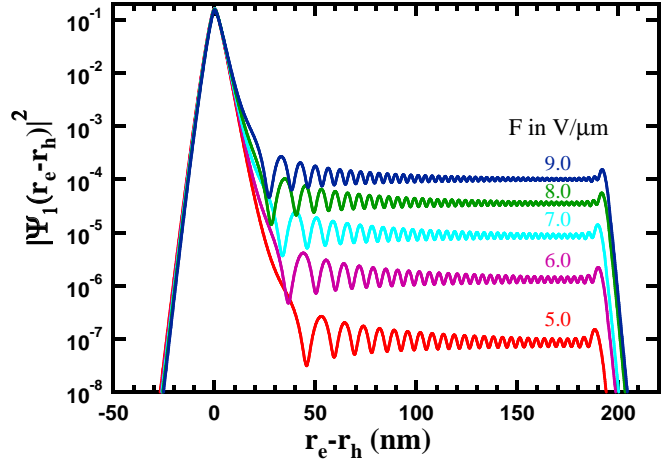}
\caption{\label{FigWavef} Optically active exciton wavefunction for
different fields $F=$ 5.0 (red), 6.0 (magenta), 7.0 (cyan), 8.0
(green), and 9.0 (blue) V/$\mu$m. There is a finite probability for
exciton to dissociate, which is found from the wavefunction weight
outside the central region $\vert r_e-r_h \vert \ge 50$ nm.}
\end{figure}

Under an applied field, a bound exciton can dissociate into a free
electron and a hole in the first continuum band, which may
contribute, for example, to photoconductivity \cite{Mohite}. The
exciton field dissociation rate determines the bound exciton
lifetime and results in the broadening of the $E_{11}$ absorption
peak. At high fields, we can determine the bound exciton
dissociation rate by fitting the $E_{11}$ absorption from the
Bethe-Salpeter equation solution to the Fano lineshape. The results
of the fits are shown in Fig.~\ref{FigGamma} by the black squares.
When the broadening becomes less than 1 meV, this procedure can no
longer be used. For smaller fields, we determine the lifetime by
calculating the tunneling probability of the bound exciton into the
free electron-hole continuum, from the leaking of the exciton
wavefunction away from the central region. The exciton wavefunctions
for different fields are shown in Fig.~\ref{FigWavef}. We calculate
the tunneling probability from the wavefunction weight at an
electron-hole separation larger than $50$ nm\cite{footnote1}. To
convert the tunneling probability into the decay rate, we need to
know the ``attempt'' frequency for tunneling. From the Heisenberg
uncertainty principle, an averaged momentum of the bound exciton is
$p=2\pi\hbar/\lambda$, where $\lambda$ is the exciton size.
According to the virial theorem, the exciton binding energy $E_b$ is
proportional to the kinetic energy $E_k\approx pv/2$, where $v$ is
an averaged velocity. In 3D case, $E_b=E_k$, and, therefore, one
would expect, in the case of nanotubes, an attempt frequency
$v/(2\lambda)$ to scale with $E_b/(2\pi\hbar)$. By adjusting the
attempt frequency to agree with the lifetime calculations from the
absorption linewidth, we can obtain the bound exciton dissociation
rate in a wide range of electric fields shown in
Fig.~\ref{FigGamma}. The attempt frequency varies from 1 to 1.5
times $E_b/(2\pi\hbar)$. The exciton dissociation rate can be
well-fitted by the following equation:
\begin{eqnarray}
\Gamma_0(F)=\alpha E_b{\frac{F_{0}}{F}}
\exp\left(-\frac{F_{0}}{F}\right)
 \label{eq4}
\end{eqnarray}
where $F_0=\beta E_b^{3/2}m^{1/2}/e\hbar$, $m$ is a reduced exciton
mass ($m^{-1}=m_e^{-1}+m_h^{-1}$). This form is motivated by the
solution of the hydrogen atom in an applied electric field
\cite{WKB}.  It should be noted here, that the electron and hole
dispersions in CNTs are hyperbolic, unlike parabolic dispersion
typically used in exciton dissociation description in solids. The
best fit for all 13 tubes, studied here, is obtained for
$\alpha=4.1$ and $\beta=1.74$. As seen from Fig.~\ref{FigGamma}, at
high fields $F/F_{0}\ge0.14$, the tunneling rate from
Eq.~(\ref{eq1}) deviates from the numerical result.  The criterion
for the full exciton dissociation\cite{footnote2} is satisfied at a
field strength of $F_c\approx F_{0}/2$. It is instructive to compare
the present results with those obtained for the 2D exciton
ionization rates in fields parallel to the quantum well plane.
\cite{Miller3}. The exciton ionization in the latter is one order of
magnitude smaller due to the difference in exciton mass and binding
energy.

It has recently been shown that under high bias the energetic
carriers in nanotubes excite optical phonons
\cite{Yao,Javey,Park,Perebeinos3} and generate a non-equilibrium
phonon distribution \cite{Lazzeri,Dai}, particularly when energy
dissipation to the substrate is suppressed as in suspended CNTs
\cite{Dai}. Therefore, one may expect that the produced excitons,
for example, in electroluminescence experiments \cite{Chen}, can
experience a phonon assisted dissociation. To explore this
mechanism, we need to model the energy distribution function of
excitons, which is determined by the coupling to the optical phonon
 bath and the spectral function of the pumping source of excitons.
The optical phonon  bath is assumed to be at temperature $T_{op}$
and the pumping source of excitons is assumed to be tuned at the
$E_{11}$ energy with intensity $P_0$. The bound exciton with zero
momentum can absorb an optical phonon with momentum $-q$ and be
promoted to state $q$ with an optical phonon energy
$\hbar\omega_{ph}$ above the bottom of the exciton band $E_{11}$.
The probability of this event is proportional to the optical phonon
occupation number $n_{ph}(-q)$ and the strength of the
exciton-optical phonon coupling. In the reverse process, the high
energy $q$-exciton can emit a phonon and be relaxed to the bottom of
the exciton band. This probability is proportional to the product of
$(1+n_{ph}(-q))$ and the strength of the exciton-optical phonon
coupling. Since the phonon dispersion is much smaller than the
exciton dispersion, we can reduce the problem to the two level
system described by the rate equations:
\begin{eqnarray}
\frac{\partial N_0}{dt}&=&P_0-
N_0\frac{\Gamma_0}{\hbar}-N_0\frac{n_{ph}}{\tau_{ph}}+N_1\frac{1+n_{ph}}{\tau_{ph}}
\nonumber \\
\frac{\partial N_1}{dt}&=&-
N_1\frac{\Gamma_1}{\hbar}-N_1\frac{1+n_{ph}}{\tau_{ph}}+N_0\frac{n_{ph}}{\tau_{ph}}
 \label{eq5}
\end{eqnarray}
where $N_0$ and $N_1$ are the ground and vibrationally excited
 states, $n_{ph}$ is given by the Bose-Einstein factor, and the
 scattering time
$\tau_{ph}\approx 30-100$ fs is determined by the exciton-optical
phonon coupling strength. The exciton-optical phonon scattering time
has been calculated \cite{Perebeinos4} and is of the same order of
magnitude with the measured \cite{Yao,Javey,Park} and calculated
\cite{Perebeinos3,Lazzeri} electron-optical phonon scattering time.
The dissociation rate of the excited state $\Gamma_1$ is much faster
than that of the ground state $\Gamma_0$. It can be modeled by
Eq.~(\ref{eq4}) with binding energy replaced by
$E_b-\hbar\omega_{ph}$, where $\hbar\omega_{ph}\simeq 200$ meV is
the optical phonon energy. Note that in the absence of the exciton
pumping source $P_0=0$ and the exciton decay channels
$\Gamma_{0,1}=0$, equation~\ref{eq5} gives a steady state thermal
exciton distribution in equilibrium with the optical phonon sink
bath. The net exciton dissociation rate, including the phonon
assisted mechanism, can be obtained from:
\begin{eqnarray}
\Gamma(T_{op})=\frac{\Gamma_0N_0+\Gamma_1N_1}{N_0+N_1}=\frac{\Gamma_0\left(1+n_{ph}+\tau_{ph}\Gamma_1/\hbar\right)
+\Gamma_1n_{ph}}{1+2n_{ph}+\tau_{ph}\Gamma_1/\hbar}
 \label{eq6}
\end{eqnarray}
where $N_0$ and $N_1$ give the steady state solution in
Eq.~(\ref{eq5}). In the limit of a very efficient thermalization,
$\tau_{ph}\Gamma_1\ll1$, the exciton ionization rate from
Eq.~\ref{eq6} is given by the thermal average of the decay rates of
the two states. On the other hand, in the opposite limit
$\tau_{ph}\Gamma_1\gg1$, which is realized, for example, when the
vibrationally excited state is above the band-to-band continuum,
$E_b\le \hbar\omega_{ph}$, the exciton dissociation rate depends
also on the exciton-optical phonon scattering time as
$\Gamma(T_{op})\approx\Gamma_0+n_{ph}/\tau_{ph}$. Even at room
temperature, the phonon assisted contribution to the dissociation
rate can be appreciable, especially at low fields, due to the strong
exciton-optical phonon interaction in carbon nanotubes. In
opto-electronic devices operating at high currents, the phonon
assisted mechanism would dominate the net exciton dissociation rate
due to the high optical phonon occupation.

In conclusion, we predict strong modulation of the absorption
spectra of carbon nanotubes by the electric field. We find a large
electric field enhancement of the band-to-band absorption, or the
Sommerfeld factor, and a large red Stark shift of the bound exciton,
especially in large diameter tubes. We find that the bound exciton
ionization rate depends strongly on field strength and can be in the
sub-picosecond range at moderately high electric fields. Finally,
optical phonon occupation, which can be high in opto-electronic
devices, dramatically increases exciton dissociation.


\begin{thebibliography}{}
\bibitem{footnote1} We find oscillations of the tunneling
probability with the size of unit cell $L$ and we choose its value
at the minimum.
\bibitem{footnote2} Here we use an equivalent definition of the
exciton dissociation field $F_c$, as the field at which the exciton
dissociation rate (linewidth) becomes comparable to the exciton
binding energy, i. e. $\Gamma_0(F_c)\approx E_b$.
\end{thebibliography}
\end{document}